\documentclass[letterpaper,twocolumn,10pt]{article}
\usepackage{geometry} 
\usepackage[utf8]{inputenc}
\usepackage[T1]{fontenc}
\usepackage{hyperref}
\usepackage{graphicx}
\usepackage{balance}
\usepackage{color}

\usepackage{tikz}
\usetikzlibrary{positioning}
\usetikzlibrary{calc}
\usetikzlibrary{arrows,shapes,backgrounds}

\usepackage[small,compact]{titlesec}
\usepackage[small]{caption}

\newcommand{\referee}[1]{#1}
\newcommand{\name}{\texttt{spark-fits}}

\newenvironment{shortitem}
{\begin{list}{$\bullet$}{\topsep=0pt\itemsep=0pt\parsep=0pt\parskip=0pt\leftmargin=12pt}}
{\end{list}}

\begin{document}

\title{FITS Data Source for Apache Spark}
\author{
{\rm Julien Peloton\footnote{\texttt{peloton@lal.in2p3.fr}}, Christian Arnault, St\'ephane Plaszczynski}
\\
LAL, Univ. Paris-Sud, CNRS/IN2P3, Universit\'e Paris-Saclay, Orsay, France
}

\maketitle

\begin{abstract}
We investigate the \referee{performance} of Apache Spark, a cluster computing framework, for analyzing data from future LSST-like galaxy surveys.
Apache Spark attempts to address big data problems have hitherto proved successful in the industry, \referee{but its use in the astronomical community still remains limited}.
\referee{We show how to manage complex binary data structures handled in astrophysics experiments such as binary tables stored in FITS files, within a distributed environment.}
To this purpose, we first designed and implemented a Spark connector to handle sets of arbitrarily large FITS files, called \name.
The user interface is such that a simple file "drag-and-drop" to a cluster gives full advantage of the framework.
We demonstrate the very high scalability of \name\ using the LSST fast simulation tool, CoLoRe, and present the methodologies for measuring and tuning the performance bottlenecks for the workloads, scaling up to terabytes of FITS data on the Cloud@VirtualData, located at Universit\'e Paris Sud. 
We also evaluate its performance on Cori, a High-Performance Computing system located at NERSC, and widely used in the scientific community.
\end{abstract}

\section{Introduction}

The volume of data recorded by current and future High Energy Physics \& Astrophysics experiments, and their complexity require a broad panel of knowledge in computer science, signal processing, statistics, and physics.
Precise analysis of those data sets is a serious computational challenge, which cannot be done without the help of state-of-the-art tools.
This requires sophisticated and robust analysis performed on many machines, as we need to process or simulate data sets several times.
Among the future experiments, the Large Synoptic Survey Telescope (LSST \cite{LSST}) will collect terabytes of data per observation night, and their efficient processing and analysis remains a major challenge.

\referee{Early 2010}, a new tool set into the landscape of cluster computing: Apache Spark \cite{sparkweb, Zaharia:2010:SCC:1863103.1863113, 180560}.
After starting as a research project at the University of California, Berkeley in 2009, it is now maintained by the Apache Software Foundation \cite{apache}, and widely used in the industry to deal with big data problems. 
Apache Spark is an open-source framework for data analysis mostly written in Scala with application programming interfaces (API) for Scala, Python, R, and Java.
It is based on the so-called MapReduce cluster computing paradigm \cite{dean2008mapreduce}, popularized by the Hadoop framework \cite{hadoop} \cite{shvachko2010hadoop} using implicit data parallelism and fault tolerance. 
In addition Spark optimizes data transfer, memory usage and communications between processes based on a graph analysis (DAG) of the tasks to perform, largely relying on functional programming.
To fully exploit the capability of cluster computing, Apache Spark relies also on a cluster manager and a distributed storage system.
Spark can interface with a wide variety of distributed storage systems, including the Hadoop Distributed File System (HDFS), Apache Cassandra \cite{cassandra}, Amazon S3 \cite{s3}, or even Lustre \cite{lustre} traditionally installed on High-Performance Computing (HPC) systems.
Although Spark is not an in-memory technology per se\footnote{Spark has pluggable connectors for different persistent storage systems but it does not have native persistence code.}, it allows users to efficiently use in-memory Least Recently Used (LRU) cache relying on disk only when the allocated memory is not sufficient, and its popularity largely comes from the fact that it can be used at interactive speeds for data mining on clusters.

\referee{When it started, usages of Apache Spark were} often limited to naively structured data formats such as Comma Separated Values (CSV), or JavaScript Object Notation (JSON) driven by the need to analyze huge data sets from monitoring applications, log files from devices or web data extraction for example. 
Such simple data structures allow a relatively easy and efficient distribution of the data across machines.
However these file formats are usually not suitable for describing complex data structures, and often lead to poor performance.
Over the past years, a handful of efficient data compression and encoding schemes has been developed by different companies such as Avro \cite{avro} or Parquet \cite{parquet}, and interfaced with Spark.
Each requires a specific implementation within the framework.

High Energy Physics \& Astrophysics experiments typically describe their data with complex data structures that are stored into structured file formats such as for example Hierarchical Data Format 5 (HDF5) \cite{hdf5web}, ROOT files \cite{rootweb}, or Flexible Image Transport System (FITS) \cite{fitsweb}.
As far as connecting the format directly to Spark is concerned, some efforts are being made to offer the possibility to load ROOT files  \cite{sparkroot}, or HDF5 in e.g. \cite{spark-hdf5, h5spark}.
For the FITS format indirect efforts are made (e.g. \cite{2011ASPC..442...93W, DBLP:journals/corr/ZhangBNSZFPP15, brahem2016astrospark}) to interface the data with Spark but no native Spark connector to distribute the data across machines is available to the best of our knowledge.

In this work, we investigate how to leverage Apache Spark to process and analyse future data sets in astronomy.
\referee{We study the question in the context of the FITS file format \cite{1981A&AS...44..363W, pence2010definition} introduced more than thirty years ago with the requirement that developments to the format must be backward compatible. This data format has been used successfully to store and manipulate the data of a wide range of astrophysical experiments over the years, hence it is considered as one of the possible data format for future surveys like LSST.}
More specifically we address a number of challenges in this paper, such as:

\begin{shortitem}
\item How do we read a FITS file in Scala (Sec. \ref{sec:sparkfits-fitsio})?
\item How do we access the data of a distributed FITS file across machines (Sec. \ref{sec:connector})?
\item What is the effect of caching the data on IO performance in Spark (Secs. \ref{sec:IO_benchmarks} \& \ref{sec:hpc})?
\item What is the behavior if there is insufficient memory across the cluster to cache the data (Sec. \ref{sec:IO_benchmarks})?
\item What is the impact of the underlying distributed file system (Sec. \ref{sec:hpc})?
\end{shortitem}
To this purpose, we designed and implemented a native Scala-based Spark connector, called \name\, to handle sets of FITS files arbitrarily large.
\name\ is an open source software released under a Apache-2.0 license and available as of April, 2018 from \url{https://github.com/astrolabsoftware/spark-fits}.

The paper is organized as follows.  We start with a review of the architecture of \name\ in Sec. \ref{sec:sparkfits}, and in particular the Scala FITSIO library to interpret FITS in Scala, the Spark connector to manipulate FITS data in a distributed environment, and the different API (Scala, Java, Python, and R) allowing users to easily access the set of tools. 
In Sec. \ref{sec:IO_benchmarks} we evaluate the performance of \name\ on a medium size cluster with various data set sizes, and we detail a few use cases using a LSST-like simulated galaxy catalogs in Sec. \ref{sec:usecase}.
We discuss the impact of using \name\ on a HPC system in Sec. \ref{sec:hpc}, and conclude in Sec. \ref{sec:conclusion}. 

\section{FITS data source for Apache Spark} \label{sec:sparkfits}

This section reviews the different building blocks of \name.
We first introduce a new library to interpret FITS data in Scala, and then detail the mechanisms of the Spark connector to manipulate FITS data in a distributed environment. 
Finally we briefly give an example using the Scala API.

\subsection{Scala FITSIO library} \label{sec:sparkfits-fitsio}

In this section we recall the main FITS format features of interest for this work.
A FITS file is logically split into independent FITS structures called Header Data Unit (HDU), as shown in the upper panel in Fig. \ref{fig:data_split}.
By convention, the first HDU is called primary, while the others (if any) are called extensions.
All HDU consist of one or more 2880-byte header blocks immediately followed by an optional sequence of associated 2880-byte data blocks.
The header is used to define the types of data and protocoles, and then the data is serialized into a more compact binary format.
The header contains a set of ASCII text characters (decimal 32 through 126) which describes the following data blocks: types and structure (1D, 2D, or 3D+), size, physical informations or comments about the data, and so on.
Then the entire array of data values are represented by a continuous stream of bytes starting at the beginning of the first data block.

\name\ is written in Scala, \referee{like Spark.}
Scala is a relatively recent general-purpose programming language (2004) providing many features of functional programming such as type inference, immutability, lazy evaluation, and pattern matching.
While several packages are available to read and write FITS files in about 20 computer languages, none is written in Scala.
There are Java packages that could have been used in this work as Scala provides language interoperability with Java, but the structures of the different available Java packages are not meant to be used in a distributed environment and they lack functional programming features.
Therefore we released within \name\ a new Scala library to manipulate FITS file, compatible with Scala versions 2.10 and 2.11\footnote{Scala versions are mainly not backwards-compatible.}.
The library is still at its infancy and only a handful of features included in the popular and sophisticated C library CFITSIO \cite{fitsio} is available.
More specifically, we first focused on providing support for reading and distributing across machines header data and binary table HDU.
\referee{At the time of writing, there is also support for image HDU (version 0.6.0) but this manuscript does not detail it.}
More features will come alongside the development of the package.

\subsection{\name\ connector} \label{sec:connector}

In Apache Spark, any complex data format needs to be described in terms of compression and decoding mechanisms.
But we also must provide the keys to distribute the computation \textit{and} access the data properly across the machines.
We need to tell Spark how to read the data from the distributed file system and describe the various operations in parallel to be sent to the worker nodes and launched by the executors (usually one executor per worker node).
Among the different distributed file systems available nowadays, this work focuses on the widely used Hadoop Distributed File System (HDFS).

\begin{figure}[!htbp]
\begin{center}
\includegraphics[width=0.5\textwidth]{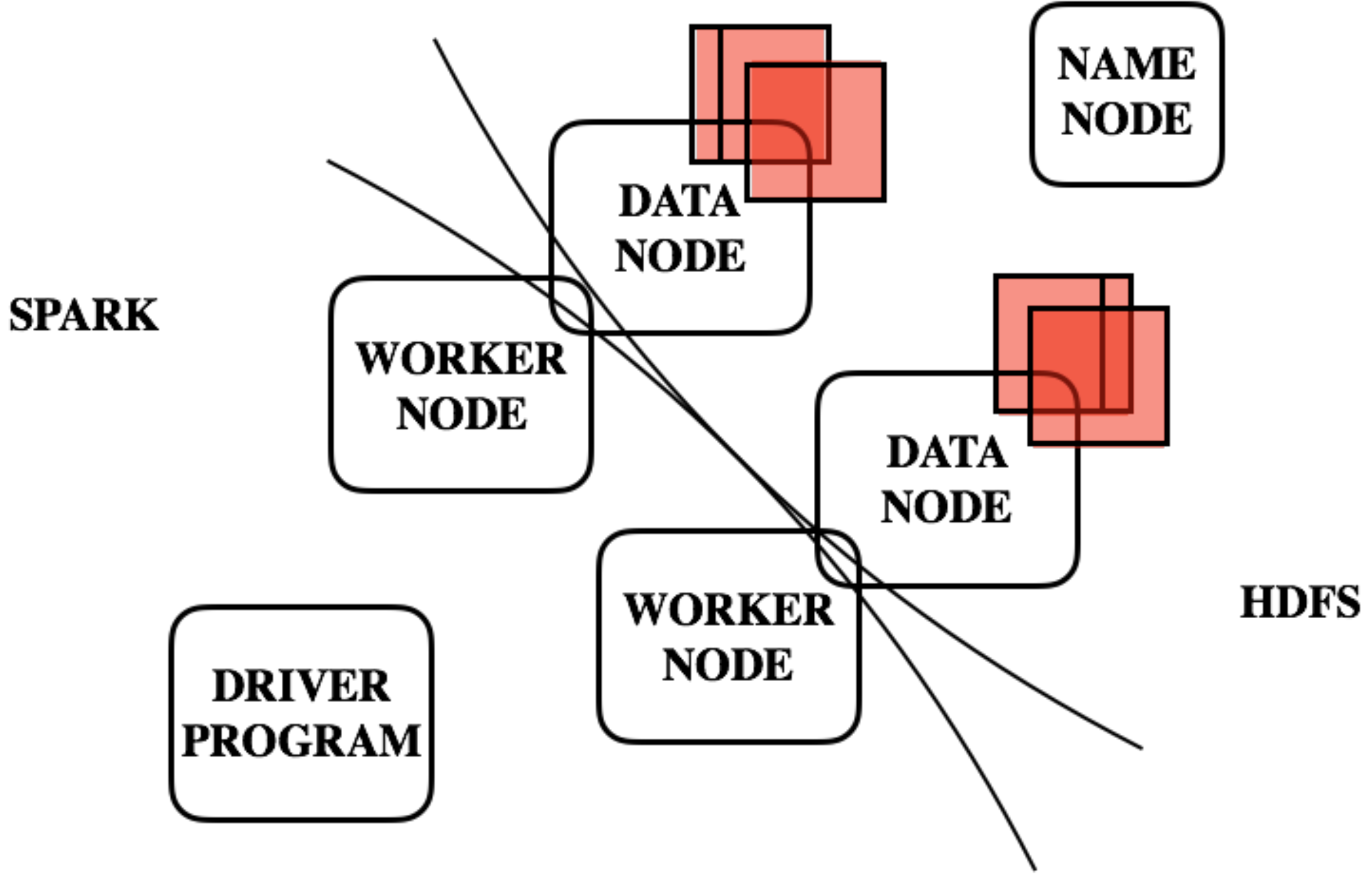}
\end{center}
\caption{On the one hand, the file blocks (red) are distributed and stored among the different DataNodes of the HDFS cluster. On the other hand, Spark's driver program, after requesting data location to the NameNode, sends computation to worker nodes as close as possible to the DataNodes where the data are stored.}
\label{fig:spark_HDFS_relation}
\end{figure}

In a HDFS cluster, an entire data file is physically split (once) into same-size file blocks which are stored inside the different DataNodes of the cluster.
A DataNode hosts only a subset of all the blocks.
Following the principle that moving computation is usually cheaper than moving data, Spark reads file blocks in a performant way: instead of copying file blocks to a central compute node, which can be expensive, the driver sends the computation to worker nodes close to DataNodes where the data reside, as shown in Fig. \ref{fig:spark_HDFS_relation}.
This is achieved after interacting with the NameNode, the master server that manages the file system namespace and regulates access to files by clients.
This procedure ensures as much as possible the principle of data locality.
Finally the distributed code processes the file blocks simultaneously, which turns out to be very efficient.
Note that for robustness, we use a replication factor of 3.

\begin{figure}[!htbp]
\begin{center}
\begin{tikzpicture}[scale=0.8,every node/.style={minimum size=1cm},on grid]

        \begin{scope}[
        yshift=-100,every node/.append style={
        yslant=0.5,xslant=-1.3},yslant=0.5,xslant=-1.3]
        
        \draw[black,very thick] (0,-0.3) rectangle (1,4);
        
        \fill[black!20!green, fill opacity=0.5] (0,1) rectangle (1,3.8);
        
        \draw[black,thin] (0,3.8) rectangle (1,4);
        \fill[black!20!green, fill opacity=0.5] (0,3.8) rectangle (1,4);
        
        \draw[-latex,thick] (1,4.1) node[right]{HEADER \# 0}
         to[out=180,in=90] (1,4.1);

	\draw[-latex,thick] (1,2.5) node[right]{DATA \# 0}
         to[out=180,in=90] (1,2.5);
         
         \draw[-latex,thick] (1,1.1) node[right]{HEADER \# 1}
         to[out=180,in=90] (1,1.1);

	\draw[-latex,thick] (1,0.3) node[right]{DATA \# 1}
         to[out=180,in=90] (1,0.3);

        \draw[black,thin] (0,0.8) rectangle (1,1);

    \end{scope}

    \begin{scope}[
        yshift=-140,every node/.append style={
        yslant=0.5,xslant=-1.3},yslant=0.5,xslant=-1.3]
        
        \draw[black,very thick] (0,-0.3) rectangle (1,4);
        
        \fill[blue, fill opacity=0.5] (0,1) rectangle (1,3.8);
         
        \draw[black,thin] (0,3.8) rectangle (1,4);
        
        \draw[black,dashed] (0,2) rectangle (1,3);

        \draw[black,thin] (0,0.8) rectangle (1,1);
        
    \end{scope}
    
    \begin{scope}[
        yshift=-180,every node/.append style={
        yslant=0.5,xslant=-1.3},yslant=0.5,xslant=-1.3]
        
        \draw[black,very thick] (0,3) rectangle (1,4);
        
        \draw[black,thin] (0,3.8) rectangle (1,4);
        \fill[black!5!red, fill opacity=0.5] (0,3) rectangle (1,4);
        
        \draw[black,very thick] (0,1.5) rectangle (1,2.5);
        \fill[black!5!red, fill opacity=0.5] (0,1.5) rectangle (1,2.5);
        
        \draw[black,very thick] (0,0) rectangle (1,1);
        \fill[black!5!red, fill opacity=0.5] (0,0) rectangle (1,1);
        \draw[black,thin] (0,0) rectangle (1,0.2);

        \draw[black,very thick] (0,-1.5) rectangle (1,-0.5);
        \fill[black!5!red, fill opacity=0.5] (0,-1.5) rectangle (1,-0.5);

    \end{scope}
\end{tikzpicture}
\end{center}
\caption{Different views of a FITS file. \textit{Upper:} The FITS format is defined in terms of structures (HDU, green), each containing a header in ASCII and binary data. \textit{Middle:} \name\ distributes the data from one HDU among executors in nearly same-length partitions (blue). \textit{Bottom:} The FITS file is physically split in same-size file blocks in HDFS regardless of its logical structure (red). Each HDFS DataNode will have a subset of all the blocks. See text for more informations.}
\label{fig:data_split}
\end{figure}
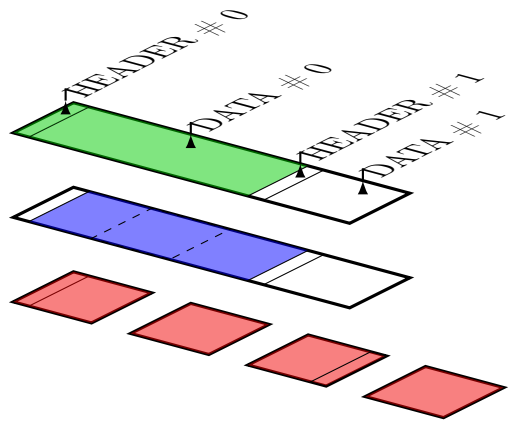

However the file block sizes in HDFS are arbitrary in the sense that HDFS does not know about the logical structure of the file format.
The file is split at regular intervals (typically a data block in HDFS has a size of 128~megabytes on disk), as shown in the lower panel in Fig. \ref{fig:data_split}.
This means for example that a file block can begin or end in the middle of a sequence of bytes or in a middle of a table row, and the remainder is contained in the next file block.

Therefore in \name\, the data is logically split into partitions following approximatively the HDFS blocks but respecting the logical structure of the file, as highlighted in the middle panel in Fig. \ref{fig:data_split}.
This logical partitioning of the data happening at the level of Spark mappers (a mapper is typically a machine core) tends to break the model of data locality. 
A mapper will probably need to read remotely (from another DataNode) some part of the missing data in its local DataNode.
However the overhead can be small if the missing data size is small compared to that of a file block or a logical partition.
In order to keep this overhead as much small as possible, each mapper loads and decodes the data as driven by its logical partition record-after-record, until the end of the partition. 
The size of the records is arbitrary and depends on what needs to be read (table rows, table columns, or image subsets for example) but it is typically smaller than the total partition size, to minimize the overhead due to data transfer between DataNodes.

\subsection{RDD and DataFrame}
At the end of the procedure, all the data from the HDU is available as a partitioned collection of records across all machines, called Resilient Distributed Datasets (RDDs) \cite{180560}.
RDDs are distributed memory abstractions, fault-tolerant and immutable.
They are designed for a general reuse (e.g. performing interactive data mining) as, once the RDDs materialized, Spark keeps them persistent in memory by default, and spill them to disk if there is not enough RAM available\footnote{Users can specify which RDDs they want to reuse and the storage strategy to execute: in-memory only, or shared between disk and memory.}.
This means that if we have to query several times the data set and enough RAM is available in the cluster, the subsequent iterations will be faster compared to the initial one.

The structuration of the files into header/data structures, and the standardisation of the format is a valuable asset.
\referee{The header of the FITS file contains all information concerning the data types therefore one can build automatically the DataFrame\footnote{A DataFrame is a distributed collection of records, like a RDD, organized into named columns and including the benefits of Spark SQL's execution engine. This is a new interface added in Spark version 1.6.} schema, hence speeding up the conversion between RDD and DataFrame.}

\subsection{Scala, Java, Python, and R API}

We released high level API for Scala, Java, Python, and R to allow users to easily access the set of tools described in the previous sections, \referee{compatible with Apache Spark version 2.0 and later}.
The API are similar to other Spark built-in file formats (e.g. CSV, JSON, Parquet, Avro).
For the sake of clarity we present only the Scala API in this section, and information about other API can be found in the documentation of \name.
Let us suppose we moved terabytes of data catalogs into a Hadoop distributed file system and we want to process it. \referee{We will use the \texttt{spark-fits} connector to connect to the data:}

\begin{verbatim}
 // Create DataFrame from the first HDU
 val df = spark.read
               .format("fits")
               .option("hdu", 1)
               .load("hdfs://...")
\end{verbatim}
No data transfer or work have been performed on the cluster, since no actions (in the sense of functional programming) have been called yet (\textit{lazyness}).
\referee{The user can then use all the DataFrame generic commands} to quickly inspect which data is contained in the HDU thanks to the header information:

\begin{verbatim}
 // The DataFrame schema is inferred from
 // the FITS header
 df.printSchema()
  root
    |-- target: string (nullable = true)
    |-- RA: float (nullable = true)
    |-- Dec: float (nullable = true)
    |-- Index: long (nullable = true)
 
 // Show the first four rows.
 df.show(4)
  +----------+---------+----------+-----+
  |    target|       RA|       Dec|Index|
  +----------+---------+----------+-----+
  |NGC0000000| 3.448297| -0.338748|    0|
  |NGC0000001| 4.493667| -1.441499|    1|
  |NGC0000002| 3.787274|  1.329837|    2|
  |NGC0000003| 3.423602| -0.294571|    3|
  +----------+---------+----------+-----+
 
\end{verbatim}
We can keep manipulating the data set by applying transformations onto it (\textit{map}, \textit{filter}, \textit{union}, ...) and only when the data exploration has reduced most of the data set size, we perform an action (\textit{count}, \textit{reduce}, \textit{collect}, ...) which triggers the computation.

There are two parts to the problem: loading of the data and computation. 
Ideally, the data will be loaded only the first time and the totality or some part of it will be kept in-memory, such that the subsequent data explorations will be limited by the computation time:

\begin{verbatim}
 // Select only 2 columns and cache
 // the data set at the next action call
 val sub_df = df.select($"RA", $"Dec")
                .persist(...)
                     
 // Perform data exploration iteratively
 for (index <- 1 to nIteration) {
   val result = sub_df.filter(...)
                      .map(...)
                      .collect()
   ...
 }
\end{verbatim}

The caching of the data will be done during the first iteration, hence cutting down the loading time for subsequent iterations.
We present the benchmarks in Sec. \ref{sec:IO_benchmarks}.

\section{Benchmarks}\label{sec:IO_benchmarks}
Unless specified, our computations are performed on a dedicated cluster installed at Universit\'e Paris-Sud (Cloud@VirtualData), France with 9 Spark executors (machines), each having 17 cores and 30 gigabytes (GB) of RAM total.
\referee{The infrastructure follows the description provided in Sec. \ref{sec:connector}, with one executor per worker node.}
The amount of memory dedicated to the caching of data is 162~GB across the cluster\footnote{We set the total memory fraction dedicated to the caching to 0.6, which corresponds to 18~GB of RAM total per executor.}.
We use a HDFS cluster for storage with 9 DataNotes, each of 3.9 terabytes (TB), and with 128 megabytes (MB) file block size.
The data sets consist in FITS files generated by the simulation tool CoLoRe \cite{colore}, and containing binary table HDU \referee{described in App. \ref{app:bench}}.
Before each test, the OS buffer caches are cleared to measure IO performance accurately.

\begin{figure}[!htbp]
\begin{center}
\includegraphics[width=0.45\textwidth]{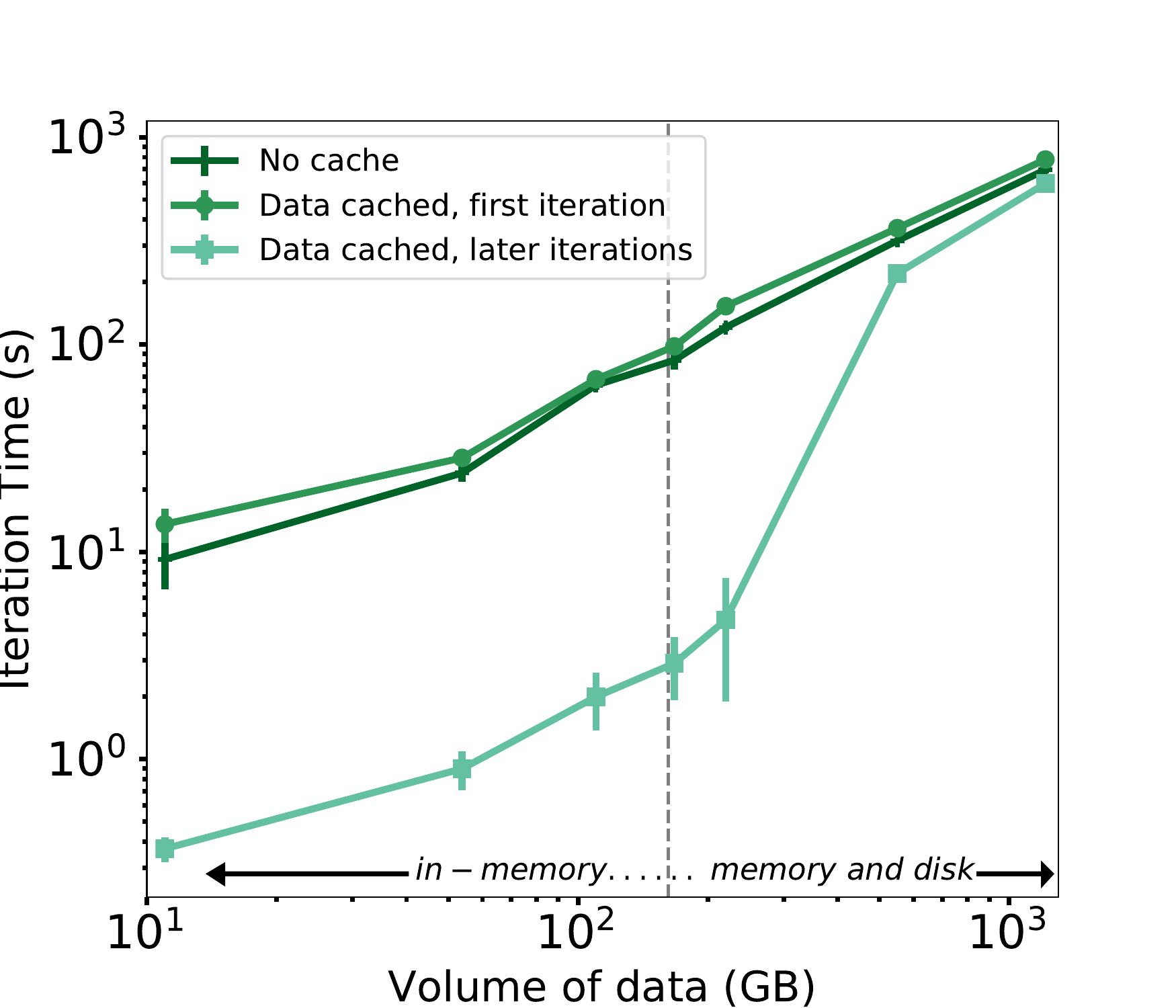}
\end{center}
\caption{Duration of iterations in Spark to load and read various FITS data sets from HDFS on 9 machines using \name. The running time at each iteration to read the data sets entirely from the disk (no caching) is shown in dark green. The iteration durations when caching the data are represented with lightest green curves: circle marks for the first iteration (loading and caching), and square marks for the later iterations. The vertical dotted line represents the maximum amount of cache memory. Error bars show standard deviations from 10 runs.
}
\label{fig:benchmark_io_lines}
\end{figure}

In order to assess the IO performance of \name, we first implement a simple application to distribute, decode, and load FITS data, and count the total number of table rows in the data sets. 
We run several times the same operations, \referee{and at the first iteration we decide if we keep the data in-memory or not.}
The results are shown in Fig. \ref{fig:benchmark_io_lines} for various data set sizes, and different strategies.

If no data is kept in memory, that is at each iteration \name\ creates partitions by reading the entire data set from disks and decoding it as described in Sec. \ref{sec:connector}, then all iterations over the data set perform at the same speed (dark green curve in Fig. \ref{fig:benchmark_io_lines}). 
We note however that the running time increases quasi-linearly with the data set size\footnote{A break of the data locality would provoke a departure form linearity.}, ranging from 11~GB to 1.2~TB.
For a fixed cluster configuration (number of executors, cores per executor, memory per core and so on), this linearity is expected due to the fact that simultaneously we have more partitions than Spark mappers, partitions are independent, and each mapper processes one partition at a time.
We measure the throughput per mapper to be around 15~MB/s (median), which corresponds to a IO bandwidth of 2.3~GB/s for the cluster.
The throughput consists of the rate to transfer data from disk (not only transferring data from a local DataNode to the worker node, but transferring data from remote DataNodes as well), and the rate to decode the data.
After analyzing the durations for several job configurations, we find that the transfer step takes about 60\% of the time, while the decoding step is about 30\%.
We are currently trying to reduce both contributions to improve the throughput.

If we want to benefit from the Spark in-memory advantages, then we have to split the work into two phases: the first iteration over the data set, and the subsequent iterations.
At the first iteration \name\ creates partitions by loading, reading and decoding the data from the disk.
In addition, partitions are put in cache by Spark until the dedicated memory fraction becomes full.
The caching operation adds an extra overhead to the running time (around 20\%), shown in circle marks in Fig \ref{fig:benchmark_io_lines}.
Once this first phase ends, the later iterations are faster to perform since all or part of the data is present in cache (that is as close as possible to the computation).
If all partitions fit into memory, then the later iteration times are almost two orders of magnitudes smaller than the first iteration time.
This is the case for the data sets up to 110~GB in our runs, as shown in square marks in Fig. \ref{fig:benchmark_io_lines}.
These performances can be used to explore the data sets at interactive speed.
If the data set size is bigger than the available cache memory, the user needs to decide whether remaining partitions will be spill to disk at the first iteration\footnote{We note also that if the initial data set is bigger than the available memory in our cluster, we can decide to cache only subsets of the data (the most re-used for example).}.
In these runs, we decide not to spill to disk remaining partitions, hence the later iterations have to recompute these partitions from scratch (transferring and decoding).
Unavoidably, the iteration time increases but the performance degrades gracefully with smaller cache fraction, making \name\ a robust tool against large data set sizes.

\begin{figure}[!htbp]
\begin{center}
\includegraphics[width=0.5\textwidth]{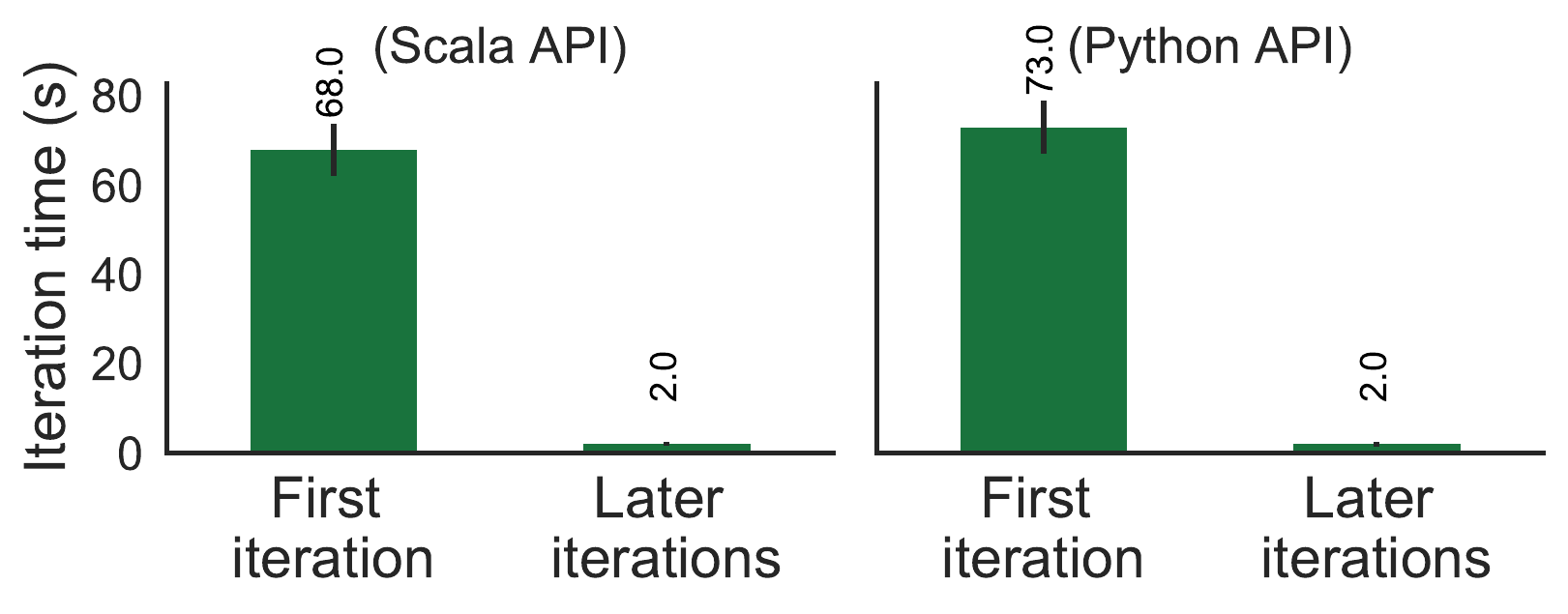}
\end{center}
\caption{Duration of the first and later iterations in Spark to load and read a 110~GB data set from HDFS on 9 machines, using the Scala API (left) or the Python API (right). The first iteration includes the time to cache the data into memory. Although Scala performs slightly better at all iterations, the performances from the two API remain comparable. Error bars show standard deviations from 10 runs.}
\label{fig:benchmark_apis}
\end{figure}

Finally we compare the performance of using the Scala API and the Python API.
The results for loading and reading a 110~GB data set are shown in Fig. \ref{fig:benchmark_apis}.
We found that using the Scala API gives slightly better performances at all iterations, but the difference remains small such that the two API can be used indifferently depending on the user needs.
Since the introduction of DataFrames in the latest versions of Spark, similar performances across API are expected.

\section{Application on galaxy catalogs} \label{sec:usecase}

We evaluated \name\ through a series of typical use cases from astronomy involving galaxy catalogs on our cluster, \referee{and described in App. \ref{app:bench}}.
Unless specified the data set used in the following experiments is made of 33 FITS files, for a total data volume of 110~GB ($6 \times 10^9$ catalog objects).
In the following experiments we make use of the Hierarchical Equal Area and iso-Latitude Pixelation, HEALPix \cite{healpix, Gorski:2004by} and more specifically the Java package, for the discretisation of the sphere, the projection in sky maps, and functions on the sphere.

\paragraph{IO benchmark.} Simply load and read the data, with no further operations. It corresponds to the benchmark described in Sec. \ref{sec:IO_benchmarks}.

\paragraph{Data split in redshift shells.}  We evaluate the cost of splitting first the catalogs data according to their redshift distance, and then projecting the data in separate sky maps (4 shells of width 0.1 in redshift distance).

\paragraph{Iterative search for neighbours in the sky.} We first define a circular region of radius 1 degree centred on a location of interest.
We then perform a coordinate transformation from RA/Dec in our set of catalogs to the corresponding HEALPix pixel index on the sky, and then we search for pixels belonging to the region of interest. We finally reduce the catalog objects found.

\paragraph{Cross-match of sets of catalogs.} We look for objects in common between two set of catalogs with size 110~GB and 3.5~GB, respectively\footnote{This corresponds to $6 \times 10^9$ and $0.4 \times 10^9$ objects, respectively.}. The cross-match is done according to the HEALPix pixel index of each object. The pixel size has been set to approximately 6.9~arcmin.  This test includes a large amount of data shuffle, at all iterations.

\begin{figure}[!htbp]
\begin{center}
\includegraphics[width=0.5\textwidth]{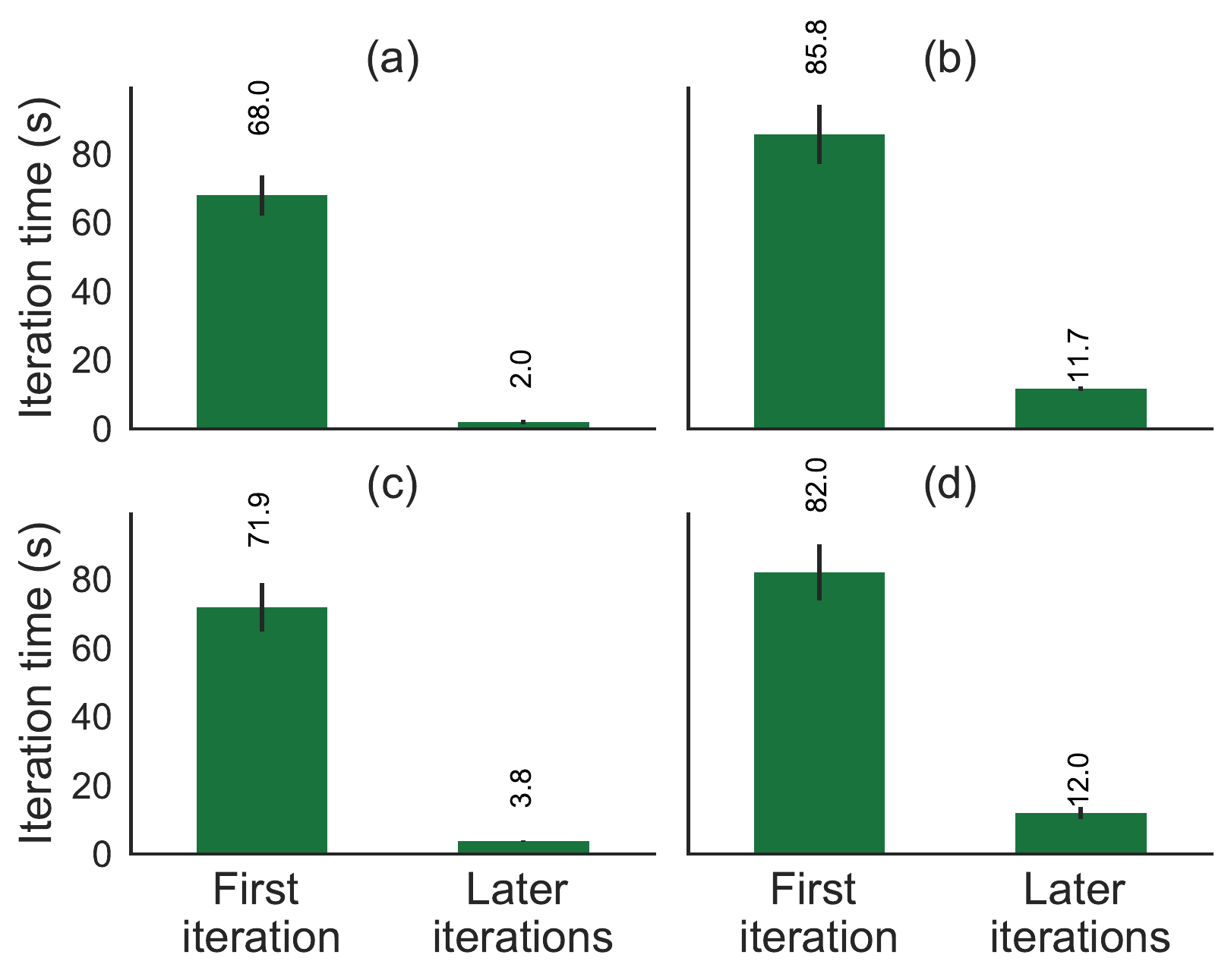}
\end{center}
\caption{Per-iteration running time of four user applications on 9 machines: IO benchmark (a), split in redshift shells (b), search for neighbours in the sky (c), and cross-match of sets of catalogs (d). The data set size is 110~GB. Error bars show standard deviations from 10 runs.}
\label{fig:benchmark_usecases}
\end{figure}

The results of the experiments are summarized in Fig. \ref{fig:benchmark_usecases}.
The first iterations are dominated by the loading and the caching of the data (case a).
For the later iterations, the running times differ but the time to read the data is often small compared to the computing time.
On overall, all the experiments ran here can be done at interactive speed once the data \referee{are} in-memory.
We plan to release a companion paper with in-depth use cases for astronomy.

\section{\name\ on HPC systems} \label{sec:hpc}

Spark can make use of HDFS, and other related distributed file systems, but is not strongly tied to those: since iterative computation can be done in memory given enough RAM, there is much less urgency in having the data local to the computation if the computation is long enough compared to the total time (including read and shuffle of the data).
Spark standalone mode and file system-agnostic approach, makes it also a candidate to process data stored in HPC-style shared file systems such as Lustre \cite{lustre}.
The use of Spark on HPC systems was explored for example in \cite{liu2016h5spark} in the context of the HDF5 data format.

We evaluate the performance of \name\ on the supercomputer Cori, a Cray XC40 system located at NERSC \cite{nersc}.
The compute nodes and the storage nodes are separated on the machine.
For the computing, the system has two different kinds of nodes: 2,388 Intel Xeon "Haswell" processor nodes and 9,688 Intel Xeon Phi "Knight's Landing" nodes. 
In this work we make use of the Haswell compute nodes, each having 32 cores and 128~GB RAM in total.
For the storage, we use the scratch file system (Lustre) with an IO bandwidth of 744~GB/sec (aggregate peak performance).
Lustre differs from block-based distributed filesystems such as HDFS in many ways.
Among those, we recall that the data is stored in a local disk filesystem. 

We used a set of 330 FITS files containing galaxy catalogs, for a total data set size on disk of 1.2~TB ($6 \times 10^{10}$ objects).
We used 40 Haswell nodes (1280 cores total), and set the number of Spark partitions as 3000.
We used Spark version 2.3.0 inside of Shifter \cite{shifter}.
In order to profile \name's performance on Cori, we ran the same benchmark as in Sec. \ref{sec:IO_benchmarks}, that is loading and reading the data and caching it at the first iteration.

\begin{figure}[!htbp]
\begin{center}
\includegraphics[width=0.5\textwidth]{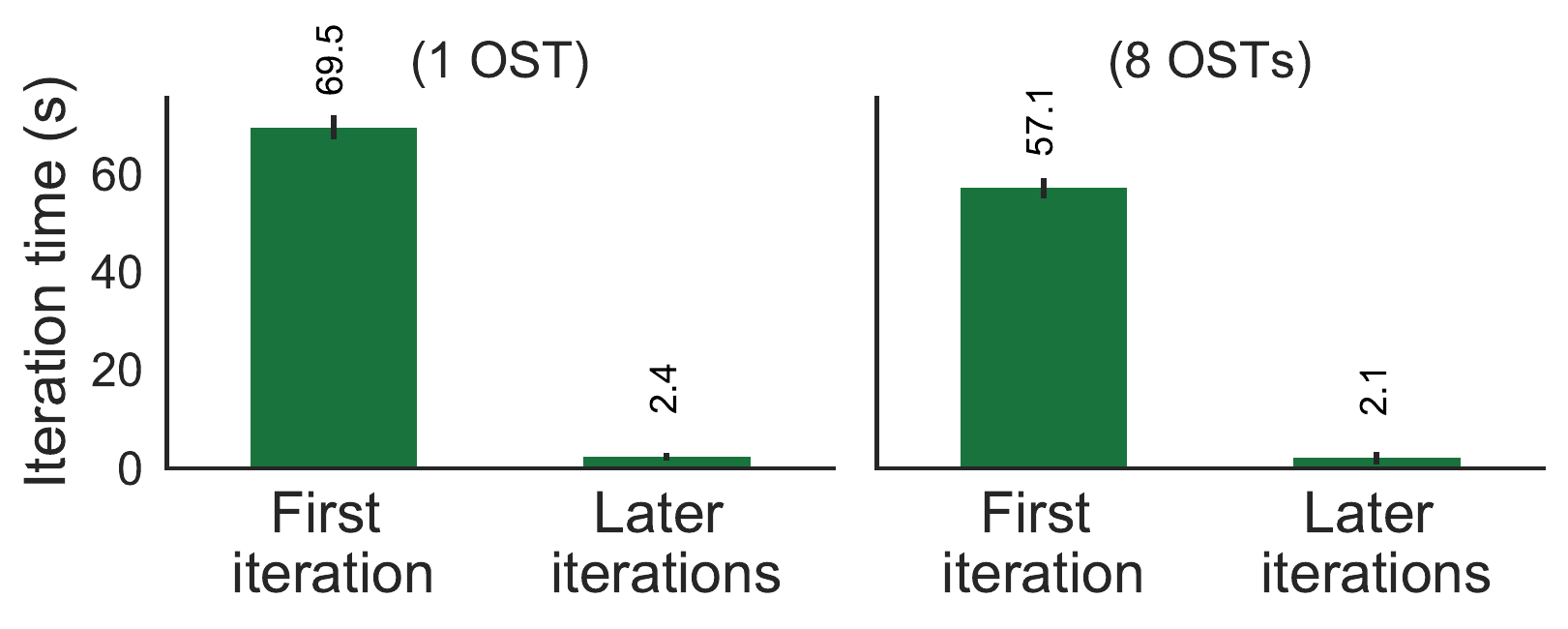}
\end{center}
\caption{Per-iteration running time to load and read a 1.2~TB data set on the supercomputer Cori at NERSC. The first iteration also includes the time for caching the data in memory. For each case we use 1280 cores, but we vary the Lustre striping count (number of OSTs): 1 (left panel) and 8 (right panel). The corresponding IO bandwidths at the first iteration are 17~GB/s and 21~GB/s, respectively. Error bars show standard deviations from 10 runs.}
\label{fig:benchmark_nersc}
\end{figure}

We show the results of the benchmark in Fig. \ref{fig:benchmark_nersc}.
The left panel presents the results using the default configuration of the system.
At the first iteration, we load the 1.2~TB of data in about a minute corresponding to a IO bandwidth of 17~GB/s.
The subsequent iterations are much faster since all the data fits into memory, and we can process the entire data set in about a few seconds (2.4 $\pm$ 0.8~s).

To increase IO performance, Lustre also implements the concept of file striping.
A file is said to be striped when read and write operations access multiple disks (Object Storage Targets, or OSTs) concurrently.
The default on Cori is a number of OST equals to 1.
We also tried a number of OST equals to 8, following the current striping recommandation on Cori for files between 1 and 10~GB.
The results are shown in the right panel of Fig. \ref{fig:benchmark_nersc}.
By increasing the number of OSTs we observe that the IO bandwidth also increases, reaching 21~GB/s.
The later iterations are also performed in about a few seconds (2.1 $\pm$ 1.3~s).
We also tried other values for striping in between 8 and 72, but we saw no significant difference with respect to the 8 OSTs case.

These results show that the use of Spark on current HPC systems gives good performance, at least equivalent to Message Passing Interface (MPI)-based softwares \cite{liu2016h5spark}.
Given enough resources, but still reasonable and accessible to scientists, one can deal with large volumes of data in a few seconds once the loading phase is done.

\section{Conclusion \& future developments} \label{sec:conclusion}

We presented \name\ a new open-source tool to distribute, read, and manipulate arbitrary large volumes of FITS data across several machines using Apache Spark.
It contains a library to interpret FITS data in Scala, and a Spark connector to manipulate FITS data in a distributed environment. 
Taking advantage of Spark, \name\ automatically handles most of the technical problems such as computation distribution or data decoding, allowing the users to focus on the data analysis.
Although \name\ is written in Scala, users can access it through several API (Scala, Python, Java, and R) without loss of performances, and the user interface is similar to other built-in Spark data sources.
At the time of writing, it is still under an active development both in optimizing the current implementation and releasing new features.
For example we are working on the possibility to distribute and manipulate image HDU which would enable a direct processing of raw data coming from telescope observations.
We note that the procedure described to deal with the FITS format in this paper can be translated with only a few minor changes to many other file formats.

We performed a series of benchmarks to evaluate the performance of \name\ against data volumes ranging from 11~GB to 1.2~TB.
For a fixed cluster configuration, the first phase which includes loading, decoding and caching the data turns out to increase linearly with the data set size.
Then assuming that enough memory is available across the cluster, the data exploration and the data analysis at the subsequent iterations can be performed at interactive speed regardless of the initial data set size.
These controlled behaviours -- linear time increase, performance as a function of cached fraction -- make \name\ a robust tool against varying data set sizes, and for a wide range of astronomical applications.
 
We also tested the use of Apache Spark and \name\ on the supercomputer Cori, located at NERSC.
We found that the use of Spark on current HPC systems gives good performance.
Using 40 Haswell compute nodes we managed to process 1.2~TB of FITS data ($6 \times 10^{10}$ objects) in a few seconds, once the loading and caching phase was done.
During the loading phase we reached an IO bandwidth of about 20~GB/s, leaving some rooms for improvement.

Our experience with \name\ indicates that cluster-computing frameworks such as Apache Spark are a compelling and complementary approach to current tools to process the expected data volumes which will be collected by future astronomical telescopes.
The intrinsic fault-tolerance of Spark and its scalability with respect to data size makes it a robust tool for many applications.
\name\ is publicly available and it is the first step towards a wider project which is aimed at building a general purpose software organization for future galaxy surveys.

 
\section*{Acknowledgements}
 
We would like to thank all Apache Spark and FITS contributors.
This research uses VirtualData infrastructures at Universit\'e Paris Sud.
We would like to thank the VirtualData team, and especially Michel Jouvin for making this work possible.
We thank C\'ecile Germain and the Moyens de Recherche Mutualis\'es (MRM) and Equipements de Recherche Mutualis\'es (ERM) for their support, and Julien Nauroy for his early contributions to the computing infrastructures.
We thank the Service Informatique at LAL, especially Adrien Ramparison and Guillaume Philippon for their technical support on the cluster, and Antoine P\'erus and Hadrien Grasland for useful discussions and comments on this work.
We would like to thank members of LSST Calcul France for valuable feedback on this work, and the anonymous referee for providing insightful comments and providing directions for additional work.
We acknowledge use of HEALPix and CoLoRe.
Some part of this research used resources of the National Energy Research Scientific Computing Center, a DOE Office of Science User Facility supported by the Office of Science of the U.S. Department of Energy under Contract No. DE-AC02-05CH11231.

On behalf of all authors, the corresponding author states that there is no conflict of interest.

\appendix

\section{Resources for benchmarks}\label{app:bench}

\referee{The data sets consist in FITS files generated by the simulation tool CoLoRe \cite{colore}, and containing binary table HDU with 5 columns (single-precision floating-point). 
The binary tables are galaxy catalogs containing celestial object coordinates (at least Right Ascension (RA) and Declination (Dec) coordinates and redshift distance information).
Each file has a size around 3.5~GB on disk.
The data are transferred (once) on our HDFS cluster prior to the runs. 

The code used to perform the IO benchmarks (Scala/Python) can be found at \url{https://github.com/astrolabsoftware/sparkioref}.
Note that this code also compares the performance of FITS with other data formats such as CSV and Parquet.
The code used to benchmark other use cases can be found at \url{https://github.com/JulienPeloton/spark-fits-apps}.
}

 \small


\end{document}